# Macroscopic Tunneling Probe of Moiré Spin Textures in Twisted CrI$_3$

Bowen Yang[1,2,†], Tarun Patel[1,2,†], Meixin Cheng[1,3], Kostyantyn Pichugin[3], Lin Tian[1,4], Nachiket Sherlekar[1,2], Shaohua Yan[5], Yang Fu[5], Shangjie Tian[5,6], Hechang Lei[5], Michael E. Reimer[1,4], Junichi Okamoto[7,8], Adam W. Tsen[1,3]*

[1]Institute for Quantum Computing, University of Waterloo, Waterloo, ON N2L 3G1, Canada.

[2]Department of Physics and Astronomy, University of Waterloo, Waterloo, ON N2L 3G1, Canada.

[3]Department of Chemistry, University of Waterloo, Waterloo, ON N2L 3G1, Canada.

[4]Department of Electrical and Computer Engineering, University of Waterloo, Waterloo, N2L 3G1, Canada.

[5]Department of Physics and Beijing Key Laboratory of Optoelectronic Functional Materials & Micro-nano Devices, Renmin University of China, 100872 Beijing, China.

[6]School of Materials Science and Engineering, Anhui University, 230601 Hefei, China

[7]Institute of Physics, University of Freiburg, Hermann-Herder-Str. 3, 79104 Freiburg, Germany.

[8]EUCOR Centre for Quantum Science and Quantum Computing, University of Freiburg, Hermann-Herder-Str. 3, 79104 Freiburg, Germany.

[†]These authors contributed equally

*Correspondence to: awtsen@uwaterloo.ca

## Abstract

Various noncollinear spin textures and magnetic phases have been predicted in twisted two-dimensional CrI$_3$ due to competing ferromagnetic (FM) and antiferromagnetic (AFM) interlayer exchange from moiré stacking — with potential spintronic applications even when the underlying material possesses a negligible Dzyaloshinskii–Moriya or dipole–dipole interaction. Recent measurements have shown evidence of coexisting FM and AFM layer order in small-twist-angle CrI$_3$ bilayers and double bilayers. Yet, the nature of the magnetic textures remains unresolved and possibilities for their manipulation and electrical readout are unexplored. Here, we use tunneling magnetoresistance to investigate the collective spin states of twisted double-bilayer CrI$_3$ under both out-of-plane and in-plane magnetic fields together with detailed micromagnetic simulations of domain dynamics based on magnetic circular dichroism. Our results capture hysteretic and anisotropic field evolutions of the magnetic states and we further uncover two distinct non-volatile spin textures (out-of-plane and in-plane domains) at ≈1° twist angle, with a different global tunneling resistance that can be switched by magnetic field.

## Introduction

There has been much interest in realizing noncollinear spin textures in magnetic systems, such as skyrmions or magnetic bubbles[1,2], and many novel proposals have been put forth to exploit such textures for next-generation spintronic information processing and memory technology[3,4]. Yet, a sizeable Dzyaloshinskii–Moriya (DM) or dipole–dipole interaction is a general prerequisite[1,2,5,6], while a substantial magnetic field may be further needed to stabilize certain spin textures[2,5]. Twisted two-dimensional (2D) or moiré heterostructures have recently emerged as a powerful new platform to introduce additional degrees of freedom to the Hamiltonian of the underlying material[7–10]. Specifically, it has been predicted that nanoscale spin textures can emerge in twisted layers of the 2D magnet CrI$_3$ even without a DM/dipole–dipole interaction or magnetic field[11,12]. These moiré textures arise naturally from the alternating ferromagnetic (FM) and antiferromagnetic (AFM) exchange coupling between adjacent layers caused by the spatially varying stacking pattern[11–14]. Although several works have reported coexisting FM and AFM layer order in twisted 2D CrI$_3$[15–19], experimental signatures of these spin textures and their field evolution are still lacking. Furthermore, it is unclear whether their states can be manipulated or readout using macroscopic electrical probes, making their technological relevance an open question.

In this work, we develop a magnetic tunneling device incorporating twisted CrI$_3$ layers and probe their spin states globally using tunneling magnetoresistance (TMR) in combination with photocurrent magnetic circular dichroism (MCD). These device-level measurements yield complementary information on the relative interlayer spin coupling and net out-of-plane magnetization, respectively. Even though the individual moiré spin textures exist on the nanoscale, our (opto-)electronic devices can sense their collective field-driven changes on the micron scale, which can then be understood via detailed micromagnetic simulations. Specifically, while MCD confirms a remnant magnetization at zero magnetic field due to magnetic domains in our devices, TMR allows us to directly distinguish between different spin textures with nearly identical net magnetization. Compared to untwisted CrI$_3$, we find that moiré samples yield memory-dependent TMR behavior owing to the additional local spin degrees of freedom. Most strikingly, the application of an in-plane magnetic field can drive a global and non-volatile transition from a ground state with out-of-plane domains to another with in-plane domains for twist angles near 1 degree. Our work demonstrates the ability to magnetically control and (opto-)electronically read out moiré spins in twisted 2D magnets using a macroscopic tunneling device, paving the way for their integration in future spintronics.

## Results

Figure 1a shows a schematic of our device and measurement geometry. As the moiré lattice in twisted bilayer $CrI_3$ (2 layers total) has been shown to exhibit significant disorder[16], we choose instead to focus on twisted double-bilayer (tDB) $CrI_3$ (4 layers total), with a small rotation angle (between 0 and 2 degrees) applied between naturally exfoliated bilayers. This twisted system exhibits a substantial FM signature that can only be attributed to the moiré stacking[15,18], as independent bilayers will always yield a layered AFM state with quenched magnetization. We sandwich the tDB $CrI_3$ between few-layer graphene (Gr) contacts and encapsulating hexagonal boron nitride (hBN) (see Methods). This geometry allows us to measure TMR in conjunction with zero-bias photocurrent ($I_{pc}$) (as well as reflection) MCD when the micron-sized junction is illuminated by a focused laser (wavelength $\lambda = 633 nm$) (see Methods, Supplementary Figure 1 for full current-voltage characteristics with and without light, and Supplementary Figure 2 for laser power dependence of $I_{pc}$). As with conventional reflection MCD, it has been recently established that photocurrent MCD (when the $I_{pc}$ difference is examined between right- and left-hand circular polarization) can be used to detect out-of-plane magnetization in untwisted $CrI_3$ junctions optoelectronically[20]. Supplementary Figure 3 shows a scanning $I_{pc}$ image taken on a 1.05 degree tDB $CrI_3$ device, which confirms that the signal is localized inside the junction area, and the main panel shows $I_{pc}$-MCD as a function of out-of-plane magnetic field (see Supplementary Figure 4 for individual polarization channels). A clear FM hysteresis loop can be seen with a marked zero field gap ($\Delta MCD$) indicating remnant magnetization, while asymmetric transitions occur near $\pm 0.8T$. These features indicate that our devices incorporate high-quality moiré $CrI_3$ with coexisting FM and AFM layer order[17,18]. Similar hysteretic behavior can be seen in both $I_{pc}$ and reflection MCD taken on another 1.05 degree device (see Supplementary Figure 5). For comparison, the inset of Fig. 1b shows $I_{pc}$-MCD on untwisted four-layer $CrI_3$ (AFM coupling only) with quenched magnetization at zero field and symmetric jumps at higher field.

The ground-state spin textures formed by the moiré pattern can be qualitatively predicted based on the $I_{pc}$-MCD results. In Fig. 1c, we plot the spatially dependent interlayer exchange coupling, $J_\perp(r)$, between the twisted layers as calculated in ref.[14] (see Supplementary Note 3 for details). As the largest FM energy (-15meV, AB stacking region) is greater than the largest AFM energy (3meV, $AB_1$' stacking region), we expect the inner twisted layers to be mostly FM-coupled with occasional AFM-coupled domains. In contrast, the outer untwisted layers always prefer AFM coupling (~0.1meV, AB' stacking). These constraints require an energy compromise as all the preferred interactions cannot be satisfied simultaneously. The left panel of Fig. 1d shows a potential resolution. Here, an out-of-plane domain is formed on one of the inner layers where the spins inside are fully flipped, becoming antiparallel (parallel) to its twisted (untwisted) neighbor. This scenario leads to a relatively low energy cost that is paid between the top two layers ($E_\perp > 0$). In comparison, the bottom two layers without domains lower the total energy for the corresponding areas ($E_\perp < 0$). The domain created also directly manifests as a noncollinear magnetic texture as the spins at the domain wall boundary are inevitably canted in-plane (see the bottom left schematic in Fig. 1d). This texture can constitute either a magnetic skyrmion or bubble depending on the winding of the in-plane spins along the domain wall[2,6].

An alternative scenario is presented in the right panel of Fig. 1d. Here, in-plane domains are formed in both inner layers; they are antiparallel but have spins pointing in-plane throughout. These domains then have a neutral interaction with their untwisted neighbors ($E_\perp \approx 0$) and generate noncollinear spins across the entire domain (see the bottom right schematic in Fig. 1d). It is not obvious *a priori* which scenario is more energetically favorable. Both have nearly identical and nonzero out-of-plane net magnetization, and thus are consistent with a finite MCD at zero field. The different spin scenarios and textures can be clearly distinguished by TMR, however, while their relative stability depends on the twist angle as we shall show. In the following section, we first perform comprehensive micromagnetic simulations of the domain dynamics as a function of magnetic field for our 1.05 degree device to elucidate the MCD findings.

Magnetic state evolution of $CrI_3$ twisted at different angles can be modeled using the anisotropic Heisenberg spin model[21] for each individual layer and constant (spatially varying) interlayer exchange coupling between the untwisted (twisted) layers along with the Zeeman interaction (see Methods and Supplementary Note 4 for details). In Fig. 2a, we plot the simulated net out-of-plane magnetization ($M_z$) of the combined 4-layer-system with 1.05 degree twist between the inner layers as a function of out-of-plane magnetic field ($B_\perp$) for sweep up (green trace) and down (orange trace). A clear FM hysteresis loop and jumps can be seen, capturing the most salient features seen from MCD in Fig. 1b and Supplementary Figure 5. In Fig. 2b, we explicitly show the simulated out-of-plane spin structure ($S_z$) in a complete moiré unit cell of each layer at several locations in the sweeps marked in Fig. 2a. Starting at a large negative field (I), we see that the layers are almost fully polarized down, except for several white regions where the spins cant in-plane. These areas correspond to strong AFM-preferred stacking between the twisted layers (see Fig. 1c). Upon increasing the field slightly (II), red (spin up) out-of-plane domains have nucleated on one of the inner layers (layer 3) at the $AB_1$' and AC' stacking regions (see Supplementary Figure 6 for comparison), thereby increasing $M_z$. Note that each of the three domains can nucleate on either of the inner layer as they are energetically equivalent. Raising the field further (III), $M_z$ exhibits a jump that corresponds effectively to a transfer of the domains to layer 2 and overall spin-reversal of the inner layers. We find this to be a general phenomenon and defer its explanation to the discussion of Fig. 3c. The spin-down domains in layer 2 become smaller when the field is brought slightly positive (IV). Note that the system retains a remnant negative $M_z$ even at zero field.

Throughout our multiple simulation runs, the spin configuration at zero magnetic field resembles the states seen in (III) and (IV). This suggests that out-of-plane domains are more likely to nucleate at 0T than in-plane domains when sweeping out-of-plane field for 1.05 degree tdB $CrI_3$. This scenario is like that shown in the left schematic of Fig. 1d. When the field is sufficiently positive (V), $M_z$ jumps up again due to spin flips of the outer layers. This $M_z$ jump at +0.8T is inherently different from the jump at -0.8T (due to domain exchange on the inner layers), and thus explains the asymmetric transitions seen in $I_{pc}$-MCD for a single sweep. The structure for the largest positive field (VI) is like (I) except all spins are reversed. Starting from this nearly fully up-polarized state and sweeping the field back slightly (VII), we find that blue (spin down) domains have now nucleated in both inner layers. Finally, when the field is reduced further and $M_z$ drops (VIII), the inner layers have again flipped overall spins and simultaneously exchanged domains, leaving a positive remnant $M_z$. Our simulations thus lend clear microscopic insight to the global magnetization dynamics of moiré magnets seen in MCD.

For potential spintronic applications, it is important to be able to detect the moiré spin textures macroscopically using purely electrical means, so we next discuss our TMR measurements on 1.05 degree tDB $CrI_3$. As we shall show, the additional local spin degrees of freedom due to the moiré coupling impart new memory-dependent TMR features compared with untwisted samples, which can again be explained by micromagnetic simulations. To demonstrate this clearly, we first show tunneling conductance as a function of $B_\perp$ for an untwisted 4-layer $CrI_3$ device for several different field sweeps (Fig. 3a). The field is initiated at -3T and ramped up to +3T for the dark green trace (only data at positive fields is shown). Starting at +3T, the field is ramped down to zero for the orange trace, and finally it is ramped up again to +3T for the light green trace. The two green traces thus correspond to the same sweep up direction and range but start at two different AFM layer configurations at 0T[22]. We can see that they overlay almost completely. The two jumps correspond to spin flips of individual layers, which are shown schematically in the insets of Fig. 3a. The orange backward sweep shows similar features except with a hysteretic shift, although there is a secondary jump at ~0.5T that can be attributed to local pinning[23,24].

In Fig. 3b and 3c, we show the equivalent measurement for 1.05 degree tDB $CrI_3$ (the dark green trace is shown in a separate panel for clarity), and qualitatively different behavior are seen for the two up sweeps. For the dark green curve (initialized at -3T), there is a sharp jump at ~1T followed by a slower increase in conductance up to 1.5T (see zoomed in inset in Fig. 3b). Here, our plot starts out of the ground state with negative $M_z$, a schematic for which is shown at the bottom left in Fig. 3b (reflecting the structure obtained from our simulations in Fig. 2b). The sharp jump corresponds to the spin flip of the outer layers (analogous to the transitions in untwisted $CrI_3$) and the smooth increase corresponds to the shrinking and gradual polarization of the domains. For the light green curve in Fig. 3c (initialized at +3T), we start out of the ground state with positive $M_z$, and increasing field does not lead to abrupt layer flipping, but rather domain changes in the inner layers, causing smooth changes in the tunneling conductance. Our simulation results for this transition are shown directly in the inset of Fig. 3c. With increasing field, the red domains in layer 2 grow into the strong FM-preferred coupling regions, which in turn causes the blue domains in layer 3 to shrink to only the AFM-preferred coupling regions. This leads to the effective transfer of domains across the inner layers and simultaneous flipping of overall spins that we have previously mentioned. The orange down sweep is the reverse of the light green transition and so the same smooth changes are observed, except with a hysteretic shift. Field-dependent photocurrent measurements and TMR taken on the other 1.05 degree device show qualitatively similar behavior (see Supplementary Figure 12).

We thus see that the presence of out-of-plane moiré domains stabilized by sweeping out-of-plane magnetic field imparts additional memory to the device behavior of 1.05 degree tDB $CrI_3$. So far the ground-state tunneling resistance at 0T appears to always have a unique value within our instrument limits, and for both positive and negative $M_z$ states. Although there is configurational entropy due to the precise placement of each domain within the inner layers, this does not change the tunneling resistance. We shall next see how this degeneracy can be lifted with the application of an in-plane field. In the main panel of Fig. 4a, we show the tunneling resistance for 1.05 degree tDB $CrI_3$ as a function of both out-of-plane magnetic field (green trace: -3T to +3T and orange trace: +3T to -3T) and in-plane field (blue trace: -7T to +7T), zoomed in around the region near 0T (see Supplementary Figure 13a for the entire sweep range). In all three cases, the resistance decreases with increasing magnetic field at higher fields as the spins become more parallel, in accordance with our expectation and simulation results in Fig. 2b. The peak resistance for the out-of-plane sweeps is offset, likely due to the hysteresis from coexisting FM order. Nonetheless, they intersect at 0T, indicating a single ground state resistance value ($R_L$) for the out-of-plane moiré lattice. Surprisingly, the blue trace clearly shows a higher tunneling resistance at 0T ($R_H$), which can only be explained by another possible spin configuration distinct from the out-of-plane domains already discussed. We note that this property is unique to tDB CrI$_3$, as untwisted 4-layers always exhibit a unique zero field resistance within the noise level (see Fig. 4a, inset). Importantly, both $R_H$ and $R_L$ states are metastable and can be switched back and forth by applying out-of-plane and in-plane fields in sequence, as shown in Fig. 4b. Measurements taken on the other 1.05 degree tDB device show qualitatively similar behavior (see Supplementary Figure 13b and 14).

To understand the nature of this spin state, we have again performed micromagnetic simulations, this time with an in-plane field. The left panel of Fig. 4c shows the ground state out-of-plane configuration obtained in Fig. 2b. Here, we have plotted $S_x$ and $S_y$ in addition to $S_z$, which can be used to visualize the domain wall windings. In this particular case, all out-of-plane domains have winding number $\pm 1$, as can be seen by their single nodes in the $S_x$ and $S_y$ plots, and so can be classified as skyrmions[2]. In general, however, we observe skyrmions with various windings ($\pm 1, \pm 2$) throughout our simulation runs. Starting from the initial configuration in Fig. 4c, we sweep an in-plane field along the x-direction up to 8.7T in our simulation and then back down. The final configuration at zero field is shown in the right panel of Fig. 4c (see Supplementary Figure 10b for the field-polarized configuration). All the out-of-plane domains are removed and in-plane domains with polarization along $\pm x$ now appear in both inner layers with AFM interlayer coupling. This scenario is like that shown in the right schematic of Fig. 1d. Our simulation results can thus be summarized as follows. After the application and removal of a polarizing out-of-plane field, a lattice with out-of-plane spins in the center of the domains is more likely to nucleate. Although these spin textures are highly metastable, subsequently applying and removing a large polarizing in-plane field is sufficient to drive a global transition and stabilize a new texture with in-plane domains.

We have further explicitly evaluated the tunneling transmission across the tDB CrI$_3$ for these various domain types (see Supplementary Note 6). Numerically, we indeed find that an in-plane domain has the largest resistance compared with all the other out-of-plane domain states. For the latter, the tunneling resistance decreases slightly with increased winding. When taking the entire moiré unit cell into account, we estimate that the low resistance state corresponding to the spin configuration on the left in Fig. 4c is ~13% smaller than the high resistance state on the right, which is comparable to that observed in our data from Fig. 4a between $R_L$ and $R_H$ (~6%). This numerical result directly substantiates our picture above. We have additionally evaluated the energies of the out-of-plane and in-plane domains and find that the difference is extremely small: ~0.03% (see Supplementary Note 6). Thus, the two textures can be considered as nearly energetically equivalent, consistent with their non-volatile nature.

The relative stability of the two domain types should be sensitive to the twist angle, however. As the moiré periodicity decreases with increasing angle, we expect out-of-plane domains where the spins must rotate by a full 180 degrees (see bottom left schematic in Fig. 1d) to become less favorable[12]. To substantiate this, we performed TMR and reflection MCD measurements on additional tDB CrI$_3$ devices with varying twist angles between 0 and 2 degrees. We can then compare the zero-field gaps observed in both measurements ($\Delta MCD$ and $\Delta R = R_H - R_L$), which is plotted in normalized percentage against twist angle in Fig. 5a (see Supplementary Figure 19 for field sweep data at larger twist angles). Distinct trends for the two measurements can be clearly observed. Near zero angle, both gaps are negligible as the system behaves mostly like untwisted CrI$_3$. As the moiré lattice emerges with increasing twist angle, both gaps grow as expected. However, while the MCD gap saturates at higher angles, the TMR gap decreases again. The former behavior is consistent with previous reports on tDB CrI$_3$ in this angle range[15], while the latter can be directly captured by our simulations. Specifically, as the twist angle is increased beyond 1.05 degrees, the nucleation of out-of-plane domains become increasingly rare—for 2 degree twist, only in-plane domains are observed for the ground state, regardless of the orientation of the initializing field (see Supplementary Figure 20). The simulated S$_z$ spin configurations for each layer at several different angles are summarized in Fig. 5b. As the net out-of-plane magnetization of both domain types are equivalent, MCD cannot resolve these differences.

## Discussion

In conclusion, through tunneling and photocurrent measurements of devices incorporating tDB CrI$_3$, we observe multiple metastable spin states that can be transitioned via the application of in-plane and/or out-of-plane magnetic fields for devices near 1 degree twist. Magnetic simulations confirm that these are due to distinct spin textures (out-of-plane and in-plane magnetic domains) with different tunneling resistance as well as their placement within the inner layers. Our findings suggest that noncollinear spins in small-twist CrI$_3$ samples may originate from either in-plane domains or out-of-plane domain walls, depending on the field conditions. Our study realizes electrical readout and magnetic control of moiré spin textures in artificially twisted 2D heterostructures. The reversible resistance switching between metastable zero-field spin textures may find potential application in novel non-volatile memory devices, while tunneling measurements with a scanning probe in place of the top graphene electrode may be used to image individual spin textures with atomic scale resolution.

## Methods

**Crystal Synthesis**

CrI$_3$ single crystals were grown by the chemical vapor transport method. CrI$_3$ polycrystals were placed inside a silica tube with 200mm length and 14mm inner diameter. After evacuation to 0.01Pa and sealing, the tubes were moved into a two-zone horizontal furnace. The temperature of the source (growth) zone was slowly raised to between 873-993K (723-823K) over a 24h period, and then held there for 150h.

**Device Fabrication**

CrI$_3$, graphite/graphene (HQ Graphene), and h-BN (HQ Graphene) were exfoliated on silicon dioxide substrates within a nitrogen-filled glovebox ($P_{O_2}, P_{H_2O} < 0.1 ppm$). Contact electrodes (17nm Au/3nm Ti) and wire bonding pads (40nm Au/5nm Ti) were pre-fabricated on sapphire and oxidized silicon wafers using photolithography and electron beam deposition. Within the glovebox, twisted double bilayer CrI$_3$ was fabricated by the tear-and-stack method[25] and the entire device heterostructure (hBN/Gr/CrI$_3$/Gr/hBN) was sequentially picked and placed on the sapphire or silicon substrate with electrodes using a polycarbonate coated polydimethylsiloxane stamp.

**Photocurrent Measurements**

Scanning photocurrent measurements were conducted in a Montana C2 cryostat at 6K. A 650nm wavelength diode laser was focused to a diffraction limited spot (~1μm) using an objective lens of *NA* = 0.55. The spot was rastered using a pair of galvo scanning mirrors and a telescopic lens system. The device photocurrent was measured using an NF Corp CA5351 current preamplifier and the reflected light was measured simultaneously using a silicon photodiode.

Photocurrent MCD measurements were conducted in an attoDRY 2100 magneto-optic cryostat at 1.6K. A 632.8nm wavelength HeNe laser, mechanically chopped at 10kHz, was focused to a diffraction limited spot (~1μm) using an objective lens of *NA* = 0.81. The laser was linearly polarized using a glan-laser polariser and subsequently circularly polarized using a quarter wave plate. The laser was parked on the sample using a white light imaging setup and x-y-z piezo stage. The photocurrent was measured using an NF Corp CA5351 current preamplifier and the voltage output was fed into an SRS 860/830 lock-in amplifier and measured at the chopping frequency.

**Reflection MCD Measurements**

Reflection MCD measurements were performed in a superconducting magnet He-4 cryostat (Cryo Industries of America) at 2K. A 632.8nm wavelength HeNe laser was focused onto the samples with a beam spot of ~2-3 μm using an aspheric condenser lens of *NA* = 0.78. The incident light was modulated between left- and right-handed circular polarization by a photo-elastic modulator (PEM; Hinds Instruments PEM-200) at 50 kHz. The reflected signal was collected by the same lens and detected by a photodiode and SRS 860/830 lock-in amplifier at 50 kHz.

**Tunneling Magnetoresistance Measurements**

Tunneling magnetoresistance measurements were performed in a superconducting magnet He-4 cryostat (Cryo Industries of America) at 2K. The devices were mounted on a single-axis rotation stage to apply magnetic fields in the directions out of plane and in plane with respect to the $CrI_3$ layers. Tunneling resistance was measured using AC voltage excitation (between 9 and 40mV) with an SRS 860 lock-in amplifier for all devices, except the untwisted 4L device shown in the main text, which was measured using a Keithley 2450 at 0.195V DC bias.

**Micromagnetic Simulations**

We simulated the magnetic structures of twisted double bilayer $CrI_3$ using a generalized Heisenberg model with exchange anisotropy and Zeeman interaction. Each layer has classical spins of *S*=3/2 on a honeycomb lattice. One moiré unit cell with periodic boundary conditions was used. Spins in adjacent layers were coupled by the interlayer exchange coupling (see Fig. 1c). The total energy was minimized by the semi-implicit method combined with backtracking line search[26] and stochastic randomness sampled from a Gaussian distribution was applied between adjacent field steps. See Supplementary Note 4 for further details.

## Data Availability

All relevant data within the article and supporting information are available from the corresponding authors upon request.

## Code Availability

The code used to simulate the spin textures of twisted double-bilayer $CrI_3$ in this study is available in the following GitHub repository: https://github.com/jo2267/Double-bilayer-CrI3-simulation. The code has also been archived on Zenodo and can be accessed by DOI:10.5281/zenodo.11241391.

## Acknowledgements

We thank Prof. Liuyan Zhao and Prof. Guo-Xing Miao for helpful discussions.

A.W.T. acknowledges support from the National Science and Engineering Research Council of Canada (ALLRP 578466-22) and the US Army Research Office (W911NF-21-2-0136). This research was undertaken thanks in part to funding from the Canada First Research Excellence Fund. J.O. acknowledges support by the G. H. Endress foundation, by the state of Baden-Württemberg through bwHPC, and by the German Research Foundation (DFG) through Grant No. INST 40/467-1 FUGG (JUSTUS cluster). H.C.L. acknowledges support from National Key R&D Program of China (Grants Nos. 2018YFE0202600 and 2022YFA1403800), Beijing Natural Science Foundation (Grant No. Z200005), National Natural Science Foundation of China (Grants Nos. 12274459), and Beijing National Laboratory for Condensed Matter Physics. The University of Waterloo's QNFCF facility was used for this work. This infrastructure would not be possible without the significant contributions of CFREF-TQT, CFI, ISED, the Ontario Ministry of Research & Innovation and Mike & Ophelia Lazaridis. Their support is gratefully acknowledged.## Author Contributions Statement

B.Y. and A.W.T. conceived and initiated the study. B.Y. fabricated the tDB $CrI_3$ devices with help from T.P. B.Y., and T.P. conducted the transport and photocurrent measurements. K.P. and M.C. designed and built the optical insert for reflection MCD, and M.C. and B.Y. performed the reflection MCD measurements. L.T., N.S., and M.R. set up and provided the attoDRY 2100 cryostat. J.O. developed the simulations. T.P. and J.O. performed the simulations with help from B.Y. S.Y., Y.F., S.T., and H.C.L. grew the $CrI_3$ crystals. B.Y., T.P., J.O., and A.W.T. wrote the manuscript with the input from all authors.

## Competing Interests Statement

The authors declare no competing interests.

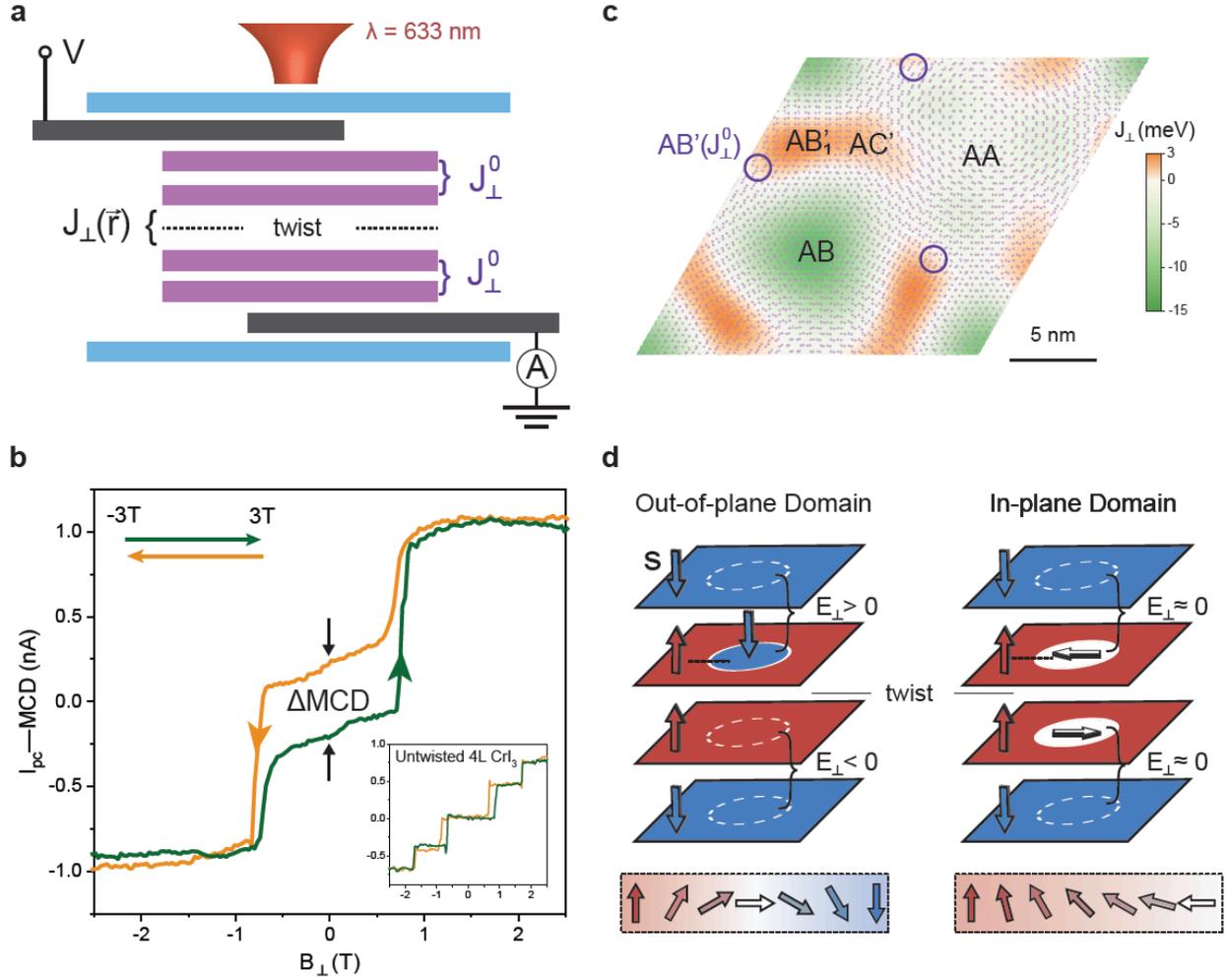

**Figure 1 | Device and (opto-)electronic measurement concept to probe moiré spin textures in tDB CrI$_3$.** (a): Sideview schematic of heterostructure device for combined tunneling magnetoresistance and photocurrent measurements under focused laser illumination. The purple, black, and blue blocks represent single-layer CrI$_3$, few-layer Gr electrodes, and few-layer hBN encapsulating flakes, respectively. The interlayer couplings, $J_\perp^0$ and $J_\perp(r)$, are labeled for untwisted and twisted CrI$_3$ layers, respectively. (b) Main panel: $I_{pc}$-MCD vs. out-of-plane magnetic field. The green (orange) curve corresponds to increasing (decreasing) field. A zero-field gap ($\Delta MCD$) is marked with two black arrows, indicating remnant magnetization. Inset: $I_{pc}$-MCD vs. out-of-plane field for untwisted four-layer CrI$_3$ for comparison. (c): Interlayer coupling map between 1.05-degree-twisted CrI$_3$ layers, calculated from ref.[14]. The atomic lattice is overlaid and several stacking regions are labeled. (d): Two possible ground-state spin textures of tDB CrI$_3$. Schematic spins across the domain walls (marked by black dashed lines) are shown below. Blue, red, and white arrows represent down, up, and in-plane spins, respectively. The sign of the interlayer coupling $E_\perp$ between the untwisted layers are evaluated for the regions with domains.

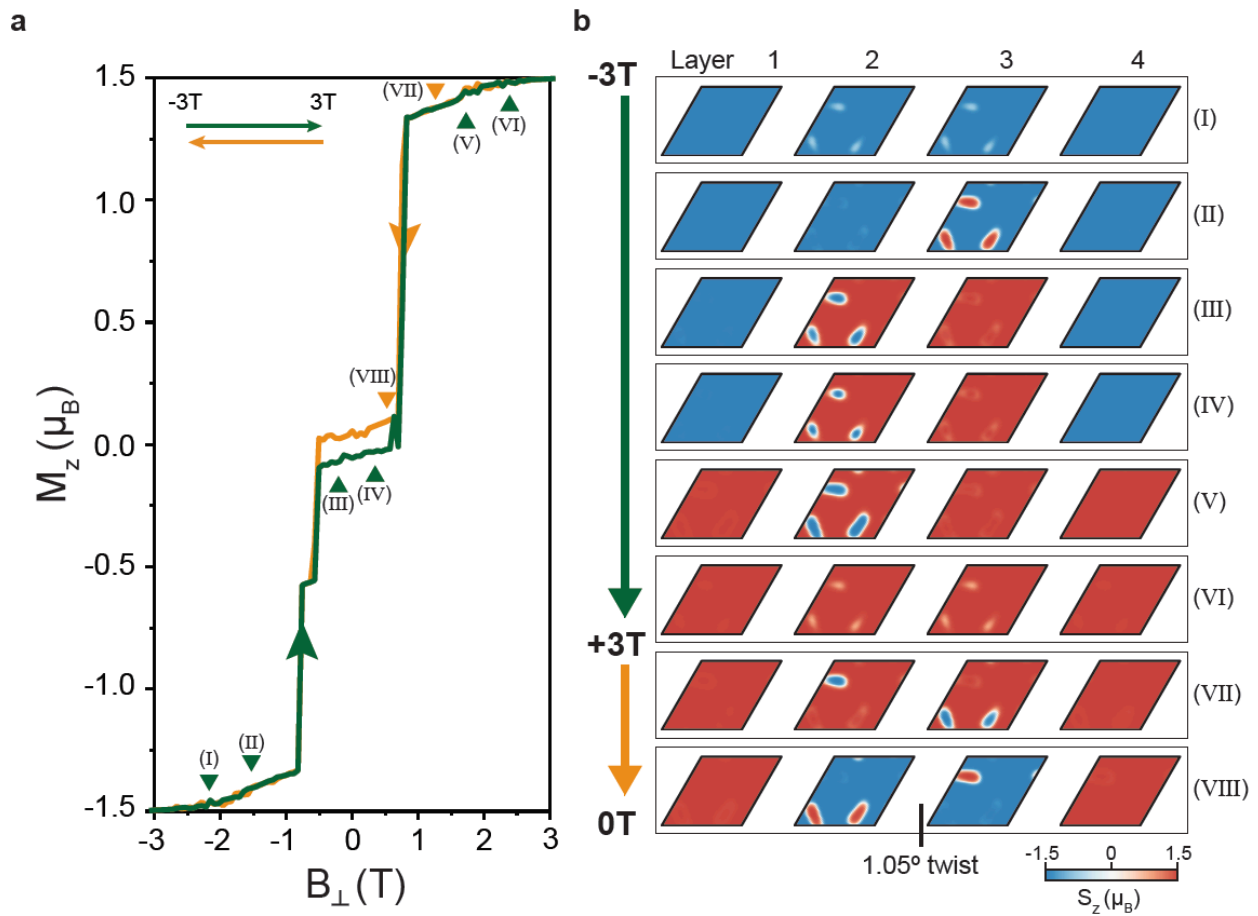

**Figure 2 | Simulated moiré spin textures under out-of-plane magnetic field.** (a): Simulated net out-of-plane component of magnetization for a moiré unit cell vs. out-of-plane magnetic field (green trace: -3T to +3T, orange trace: +3T to -3T), capturing the key experimental features in Fig. 1b. (b): Out-of-plane spin structure ($S_z$) for each CrI$_3$ layer within a moiré cell at the magnetic field positions marked in (a). The red and blue colors represent up and down spins, respectively. Out-of-plane domains are seen around zero field, consistent with the left picture in Fig 1d.

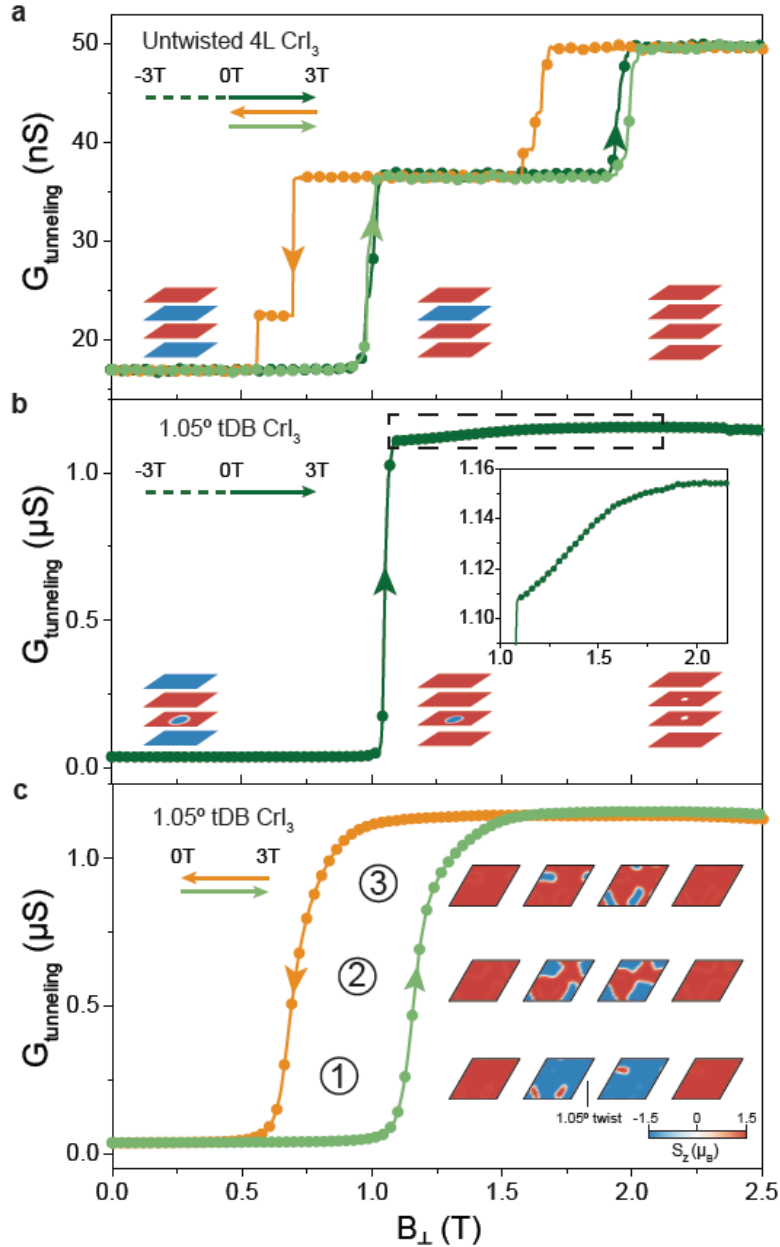

**Figure 3 | Memory-dependent tunneling magnetoresistance under out-of-plane magnetic field.** (a): Tunneling magnetoconductance for untwisted four-layer CrI$_3$ showing abrupt layer-flipping transitions regardless of sweep direction and/or starting state. Spin structures for the plateau regions are shown in the insets. (b): Tunneling magnetoconductance for tDB CrI$_3$ with field swept from -3T to +3T showing abrupt transition at ~1T corresponding to flipping of the outer layers. Schematic field-dependent spin structures centered around a single domain are shown at the bottom (red = up spins, blue = down spins, white = in-plane spins). The inset shows a zoom-in of the magnetoconductance data above the layer-flipping transition. (c): Magnetoconductance sweeps between 0T and +3T for tDB CrI$_3$ starting from a zero-field spin configuration different from that in (b). A gradual transition is observed. Insets show simulated $S_z$ for each layer across the transition. The inner layers effectively exchange domains and reverse spins.

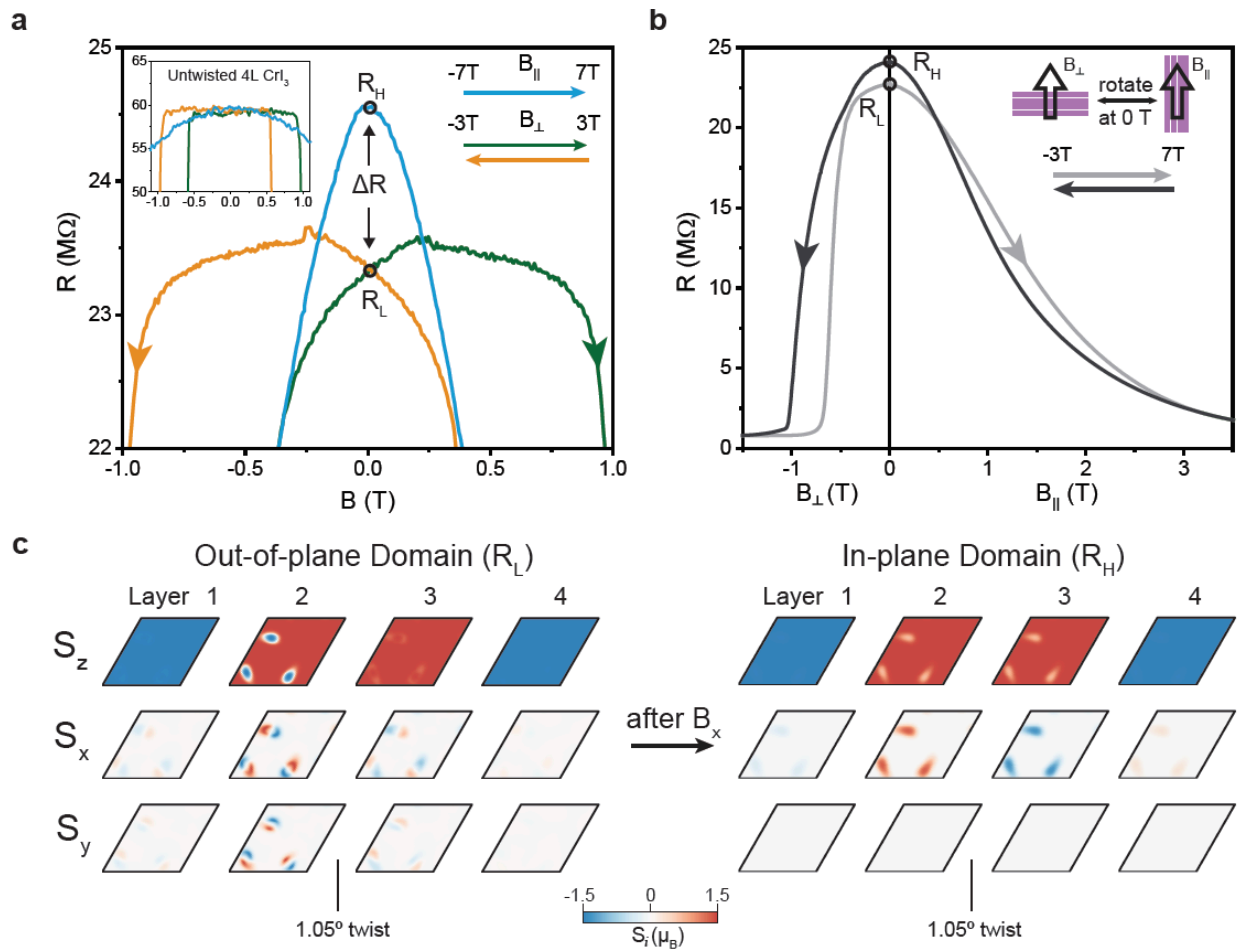

**Figure 4 | Non-volatile tunneling magnetoresistance under in-plane and out-of-plane magnetic fields from transitions between different spin textures.** (a): Tunneling magnetoresistance for tDB $CrI_3$ under out-of-plane (orange and green traces) and in-plane (blue trace) magnetic field. Two resistances at zero field are seen: $R_H$ and $R_L$. The inset shows tunneling magnetoresistance under similar conditions for untwisted $CrI_3$. A single zero-field resistance state independent of field orientation is observed within the noise level. (b): Non-volatile and reversible switching between $R_H$ and $R_L$ states realized by sequentially sweeping out-of-plane and in-plane magnetic fields. The inset shows the measurement procedure used by rotating the sample. (c): Simulated $S_z$, $S_x$ and $S_y$ spin structures before (left) and after application and subsequent removal of a polarizing in-plane field along +x (right). The system undergoes a transition between out-of-plane to in-plane domains.

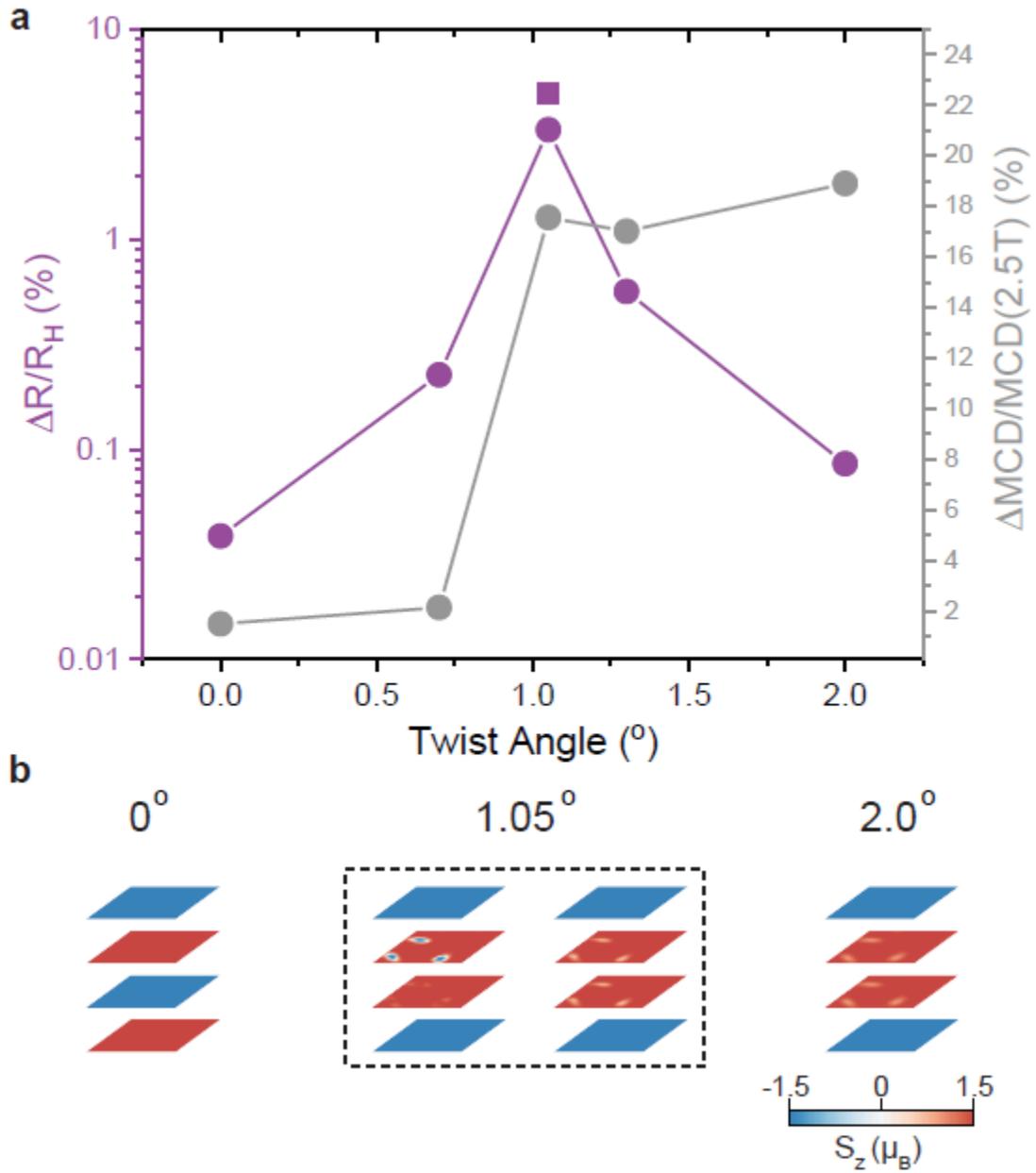

**Figure 5 | Twist angle dependence of TMR and MCD measurements.** (a): $\Delta R/R_H$ (left y-axis, purple), and reflection $\Delta MCD/MCD(2.5T)$ (right y-axis, gray) as function of twist angle (0, 0.7, 1.05, 1.3 and 2 degrees). The square data point at 1.05 deg is extracted from the TMR of Device 2 (see Supplementary Figure 13b). $\Delta MCD$ is calculated by the distance between the two linear fitting of MCD curves around zero field (see Fig. 1b). The size of the data points is larger than the error bars. (b): Micromagnetic simulations for the ground states of 0, 1.05, and 2 degree twist samples ($S_z$ channel). Both out-of-plane and in-plane domain ground states are observed at 1.05 degrees only.